\documentclass[a4paper,10pt]{article}

\usepackage{csquotes}
\usepackage[english]{babel}
\usepackage{amsfonts}

\usepackage{appendix}

\newcommand\tab[1][1cm]{\hspace*{#1}}
\usepackage{placeins}
\usepackage{pdfpages}

\usepackage{algorithm} 
\usepackage[algo2e]{algorithm2e} 
\usepackage{algpseudocode}
\algnewcommand{\LeftComment}[1]{\Statex \(\triangleright\) #1}

\usepackage{biblatex}

\usepackage{amssymb}
\usepackage{mathtools, stmaryrd}
\usepackage{xparse} \DeclarePairedDelimiterX{\Iintv}[1]{\llbracket}{\rrbracket}{\iintvargs{#1}}
\NewDocumentCommand{\iintvargs}{>{\SplitArgument{1}{,}}m}
{\iintvargsaux#1} %
\NewDocumentCommand{\iintvargsaux}{mm} {#1\mkern1.5mu..\mkern1.5mu#2}

\usepackage{amsmath}
\newcommand{\card}[1]{\left\lvert#1\right\rvert}

\DeclarePairedDelimiter\floor{\lfloor}{\rfloor}
\usepackage{graphicx}

\usepackage{enumitem}

\bibliography{references}

\usepackage{hyperref}%

\usepackage{authblk}

\newcommand{\pluseq}{\mathrel{{+}{=}}}

\date{}

\title{On the Decentralized Generation of the RSA Moduli in Multi-Party Settings}
\author[1]{Vidal Attias\thanks{vidal.attias@iota.org}}
\author[1]{Luigi Vigneri\thanks{luigxi.vigneri@iota.org}}
\author[1]{Vassil Dimitrov\thanks{vassil@iota.org}}
\affil[1]{IOTA Foundation}

\begin{document}
\maketitle

\begin{abstract}
    RSA cryptography is still widely used. Some of its applications (e.g., distributed signature schemes, cryptosystems) do not allow the RSA modulus to be generated by a centralized trusted entity. Instead, the factorization must remain unknown to all the network participants. To this date, the existing algorithms are either computationally expensive, or limited to two-party settings. In this work, we design a decentralized multi-party computation algorithm able to generate efficiently the RSA modulus.
\end{abstract}

\section{Introduction}
Rivest-Shamir-Adleman (RSA)~\cite{Rivest} is one of the first public key encryption systems, and it is still widely used. RSA uses the product of two large prime numbers $p$ and $q$ to compute a public modulus $N$. The security of RSA is guaranteed by the difficulty of the integer factorization of the public modulus: its factorization would provide the prime numbers $p$ and $q$ and, thus, the corresponding private key.

The simplest technique to generate RSA keys is by using an agent trusted by the parties who does not disclose the factorization. However, many applications using RSA (e.g., distributed signature schemes \cite{Preneel2000}, threshold cryptosystems \cite{Hazay}, multi party computations \cite{Cramer}) require to generate the RSA keys in a distributed way. In other contexts, for instance when using verifiable delay functions \cite{Boneh2018} for spam prevention in distributed ledger technologies~\cite{CoordicieTeam2019}, the private key does not have to be computed, and the algorithm require the decentralized generation of the public modulus only.

In this document, we propose a multi-party protocol generating a public RSA modulus, which extends the state-of-the-art two-party algorithm discussed in~\cite{Frederiksen}.

The rest of the paper is organized as follows: in Section~\ref{sec:state_of_art}, we present the previous work on RSA key generation, and in Section~\ref{sec:frederisken} we review in detail the fastest algorithm to date~\cite{Frederiksen}; then, in Section~\ref{sec:generalization}, we describe our algorithm working in a multi-party setting; after that, we conclude our paper with a discussion about network overhead and algorithm complexity in Section~\ref{sec:analysis}.

\section{State of the art}\label{sec:state_of_art}
Fiat and Shamir seminal paper on signature schemes~\cite{Fiat1986} has paved the way to the research on usnig public modulus with unknown factorization.
Later, Boneh and Franklin designed a distributed modulus generation algorithm including a biprimality testing algorithm in a multi-party setting~\cite{Boneh1997}. The algorithm works with semi-honest adversaries\footnote{Semi-honest nodes precisely follow the protocol (like honest nodes) but try to collect information for malicious purposes during their communications. They are also called \emph{honest but curious}.} in a honest majority. 

Another direction was taken by Algesheimer \cite{Algesheimer2002} and by Damgard and Mikkelsen \cite{Damgard2010}. The authors suggested to compute a full primality test of the factors while keeping them secret instead of a biprimality test by executing a distributed Miller-Rabin test \cite{Rabin1980}. Gilbo presented a secure version of the protocol based on the Boneh and Franklin technique \cite{Boneh1997}.

The first work on two-parties protocol with one of the party being semi-honest  has began with \cite{Cocks} but was found to be insecure by \cite{Blackburn1999}

More recently, Frederiksen \cite{Frederiksen} proposed an algorithm using a new distributed product routine running in a two-party setting both in a semi-honest and in a malicious environment by using modern techniques such as Oblivious Transfer extensions \cite{Asharov2017}; this solution runs in 40 seconds, compared to 15 minutes of the previous fastest protocol. In this paper, we propose a multi party algorithm for multi-party RSA modulus generation by extending the Frederiksen's protocol~\cite{Frederiksen} to more than two parties. We start by reviewing the original protocol in the next section.

\section{Frederiksen's protocol}\label{sec:frederisken}

\subsection{Algorithm set up}

\subsubsection{Objective}

The algorithm aims to generate two large \textit{prime} random numbers $p$ and $q$ of the same order (i.e., having the same bit-length) and the distributed computation of their product $N=p\cdot q$. The numbers $p$ and $q$ are $k$-bit numbers, i.e., $p, q \in [1, 2^k]$. A fundamental requirement is that none of the party involved in the computation can be able to retrieve either $p$ or $q$.

\subsubsection{System model}

Consider a network made of a set $\mathcal{N}$ of two nodes $\textrm{\textbf{P}}_1, \textrm{\textbf{P}}_2\in\mathcal{N}$ which decide to collaborate to run the protocol. The protocol makes the following additional assumptions:

\begin{enumerate}[label={A}.{{\arabic*}}]
    \item \textit{Semi-honest participants}. Both nodes follow the protocol. However, one of them could try to guess the secret shares of the other one to retrieve the factorization of the public modulus.
    
    \item \textit{Reliable networking layer}. Every message is received in the same way it has been sent with probability one and in finite time.
\end{enumerate}

\subsubsection{Protocol outline}

At a high level, the proposed algorithm involves the following steps:

\begin{enumerate}[label={P}.{{\arabic*}}]
    \item \textit{Random number generation}. The two active nodes $\textrm{\textbf{P}}_1$ and $\textrm{\textbf{P}}_2$ respectively generate the numbers $p_1$, $q_1$ and $p_2$, $q_2$.
    \item \textit{Fast trial divisions}. The active nodes run fast trial divisions by small primes numbers on both $p = p_1+p_2$ and $q = q_1+q_2$ to quickly discard wrong candidates. If $p$ or $q$ fail the test, the algorithm restarts from P.1.
    \item \textit{RSA modulus computation}. The two parties compute $N=(p_1+p_2)\cdot(q_1+q_2)$ in a distributed way without revealing information on key's parts.
    \item \textit{Biprimality test}. A biprimality test verifies whether $N$ is a product of two prime numbers \textit{whp}. If the biprimality test fails, the algorithm restarts from P.1.
\end{enumerate}

In the next subsection, we will describe rigorously each part of the protocol.

\subsection{Detailed protocol}

\subsubsection{Random number generation}
The goal of this step, is for node $\textrm{\textbf{P}}_1$ (resp. $\textrm{\textbf{P}}_2$) to pick two numbers $p_1$, $q_1 \in [0, 2^k-1]$ (resp. $p_2$, $q_2 \in [0, 2^k-1]$). Let $p \triangleq p_1+p_2$ and $q\triangleq q_1+q_2$ be the sum of the number generated by the two nodes. However the biprimality test (step P.4 of the protocol) requires
\begin{equation}\label{eq:biprimality_condition}
    p\equiv3(\textrm{mod}4) \ \wedge \ q\equiv3(\textrm{mod}4).
\end{equation}

Hence, the protocol enforces that $\textrm{\textbf{P}}_1$ (resp. $\textrm{\textbf{P}}_2$) randomly picks two numbers $\hat{p}_1$, $\hat{q}_1 \in [1, 2^{k-2}]$ (resp. $\hat{p}_2$, $\hat{q}_2 \in [1, 2^{k-2}]$). After that, node $\textrm{\textbf{P}}_1$ (resp. node $\textrm{\textbf{P}}_2$) concatenates two zeros (resp. two ones) to satisfy Eq.~\eqref{eq:biprimality_condition}, i.e., $p_1 \triangleq \hat{p}_1 || \{0;0\}$ and $q_1 \triangleq \hat{q}_1 || \{0;0\}$ (resp. $p_2 \triangleq \hat{p}_2 || \{1;1\}$ and $q_2 \triangleq \hat{q}_2 || \{1;1\}$).

\subsubsection{Fast trial divisions}
As a protocol optimization, we can discard some trivial non-prime candidates to reduce the number of the algorithm iterations. The idea is to divide the candidates with small prime numbers.

Let $B\in\mathbb{N}$ be a certain threshold, which indicates the number of prime numbers that we want to test. For each prime $\beta < B$, $\textrm{\textbf{P}}_1$ and $\textrm{\textbf{P}}_2$ run a $\beta$-divisibility test for $p$ and $q$. The $\beta$-divisibility test consists in computing the remainder of $(p_1+p_2)(\textrm{mod}\ \beta)$ while keeping secret $p_1$ and $p_2$. The above is described in \autoref{alg:fast_trial}. The parties test if $p_1+p_2\equiv 0\ (\textrm{mod}\ \beta)$ or, equivalently, $p_1\equiv-p2\ (\textrm{mod}\ \beta)$. We use the $1$-out-of-$\beta$ Oblivious Transfer (OT) algorithm\footnote{We refer the interested reader to Appendix~\ref{appendix:ot} for further information on the algorithm.} to hide the rest of the modulus $\beta$ of $p_2$ to $\textrm{\textbf{P}}_1$. The same test has to be ran for $q$.

In the algorithm, $\textrm{\textbf{P}}_1$ inputs $\{r_0,\dots,r_{\beta-1}\}\in[0,2^k-1]^\beta$ to the OT, while $\textrm{\textbf{P}}_2$ requests $r_{-p_2(\textrm{mod}\ \beta)}$ from the OT and send it to $\textrm{\textbf{P}}_1$. Then, $\textrm{\textbf{P}}_1$ can check if $r_{-p_2(\textrm{mod}\ \beta)} = r_{p_1(\textrm{mod}\ \beta)}$: if yes, then $p1\equiv -p_2 (\textrm{mod} \beta)$ and the protocol has to be restarted.%; otherwise, none of the parties learned more information on the secret share of the other one than their secret share doesn't have the same remainder modulus $\beta$ which is a very weak leakage of information.

Finally, one can note that this step is parallelizable by running different $\beta$-divisibility tests simultaneously with different values of $\beta$.

{\centering
\begin{minipage}{.8\linewidth}
  \begin{algorithm}[H]
        \SetAlgoLined
        $\textrm{\textbf{P}}_1$ and $\textrm{\textbf{P}}_2$ initialize a $1$-out-of-$\beta$ OT\\
        $\textrm{\textbf{P}}_1$ generates $\{r_0,\dots,r_{\beta-1}\}\in[0,2^k-1]^\beta$\\
        $\textrm{\textbf{P}}_1$ inputs $\{r_0,\dots,r_{\beta-1}\}$ to the OT\\
        $\textrm{\textbf{P}}_2$ computes $a=-p_2\mod\beta$\\
        $\textrm{\textbf{P}}_2$ receives $r_a$ from the OT\\
        $\textrm{\textbf{P}}_1$ computes $b=p_1\mod\beta$\\
        $\textrm{\textbf{P}}_1$ sends $r_b$ to $\textrm{\textbf{P}}_2$\\
        \textbf{if} $r_a=r_b$ \textbf{then} $\textrm{\textbf{P}}_2$ sends $\bot$
        else $\textrm{\textbf{P}}_2$ sends $\top$
        \caption{$\beta$-divisibilty-test}\label{alg:fast_trial}
        \end{algorithm}
\end{minipage}
\par
\vspace{5mm}
}

\subsubsection{RSA modulus computation}
This step aims to compute the product between $p$ and $q$, i.e.,
\begin{align*}
    N &= p\cdot q\\
    &=(p_1+p_2)\cdot(q_1+q_2)\\
    &=p_1\cdot q_1 + p_1\cdot q_2 + p_2\cdot q_1 + p_2\cdot q_2.
\end{align*}

While the terms $p_1\cdot q_1$ and $p_2\cdot q_2$ can be computed locally by $\textrm{\textbf{P}}_1$ and $\textrm{\textbf{P}}_2$, the two remaining terms must be computed in a distributed way. To this end, the protocol defines the procedure \texttt{distr\_product}. 

Let us assume $\textrm{\textbf{P}}_1$ knows $a\in\mathbb{N}$ and $\textrm{\textbf{P}}_2$ knows $b\in\mathbb{N}$, and the two nodes want to compute $a\cdot b$ without disclosing their number to the other party. Moreover, let
\begin{equation*}
    l\triangleq\max(\floor{\log_2(a)}+1, \floor{\log_2(b)}+1)
\end{equation*}
be the minimum number of bits necessary to represent $a$ and $b$. Then, given an input $x\in\mathbb{N}$, for bit $i$, we can define the function
\begin{center}
    \begin{math}
        \mathcal{B}_i(x) = \begin{cases} 0, & \mbox{if the i-th bit of } x \mbox{ is 0;} \\ 1, & \mbox{otherwise.}\end{cases}
    \end{math}
\end{center}
Note that we count bits from the least significant digit.

For each bit $i$ of $x$, $\textrm{\textbf{P}}_1$ generates
\begin{equation*}
    c_i^1=c_i^0+a \textrm{, where } tc_i^0\in[0,2^l-1],
\end{equation*}
and the instantiates an OT with $\{c_i^0, c_i^1\}$ as an input. Then, $\textrm{\textbf{P}}_2$ requests $c^{\mathcal{B}_i(b)}_i = c_i^0 + a\cdot\mathcal{B}_i(b)$. Finally, $\textrm{\textbf{P}}_1$ computes
\begin{equation*}
    s_1 = -\sum\limits_{i=0}^{l-1}c_i^0\cdot2^i,
\end{equation*}
and $\textrm{\textbf{P}}_2$ computes
\begin{equation*}
    s_2 = \sum\limits_{i=0}^{l-1}c^{\mathcal{B}_i(b)}_i\cdot2^i.
\end{equation*}
Hence,
\begin{equation*}
    s_1 + s_2 = \sum\limits_{i=0}^{l-1}\mathcal{B}_i(b)\cdot a\cdot2^i = a\cdot b.
\end{equation*}

To calculate $p_1\cdot q_2 + q_1 \cdot p_2$ without information leakage, $\textrm{\textbf{P}}_1$ will compute $s_1^p$ (resp. $s_1^q$) and $\textrm{\textbf{P}}_2$ will compute $s_2^q$ (resp. $s_2^p$) by running \texttt{distr\_product}($p_1$, $q_2$) (resp. \texttt{distr\_product}($q_1$, $p_2$)). Finally, $\textrm{\textbf{P}}_1$ will share $s_1^p+s_1^q$ with $\textrm{\textbf{P}}_2$ and $\textrm{\textbf{P}}_2$ will share $s_2^p+s_2^q$ such as
\begin{equation*}
    s_1^p+s_2^p+s_1^q+s_2^q = p\cdot q.
\end{equation*}

Importantly, the algorithm allows each party to only have partial information of the output, while their sum is the exact output. In order to dissimulate information, the parties only share the sum of all the partial product they compute.

{\centering
\begin{minipage}{.8\linewidth}
    \begin{algorithm}[H]
        \SetAlgoLined
        We have $\textbf{P}_1$ which knows $a=\sum a_i2^i$ and  $\textbf{P}_2$ which knows $b=\sum b_i2^i$ with $i\in\{0;l-1\}$ and $a_i, b_i\in \{0,1\}$\\
        $\textbf{P}_1$ sets $s_1=0$\\
        $\textbf{P}_2$ sets $s_2=0$\\
        \textbf{for} $i=0$ to $i=l-1$\\
        \tab $\textbf{P}_1$ and $\textbf{P}_2$ initialize a $1$-out-of-$2$ OT\\
        \tab $\textbf{P}_1$ generates $c_0\in\{0,2^l-1\}$\\
        \tab $\textbf{P}_1$ computes $c_1=c_0+a\mod 2^l$\\
        \tab $\textbf{P}_1$ inputs $\{c_0, c_1\}$ into the OT\\
        \tab $\textbf{P}_2$ receives $c_{b_i}$ from the OT\\
        \tab $\textbf{P}_1$ computes $s_1 = s_1 - c_02^i$\\
        \tab $\textbf{P}_2$ computes $s_2 = s_2 + c_{b_i}2^i$\\
        Both parties keep $s_1$ and $s_2$ secret for future usage
        \caption{\texttt{distr\_product}}
    \end{algorithm}
\end{minipage}
\par
\vspace{5mm}
}

{\centering
\begin{minipage}{.8\linewidth}
  \begin{algorithm}[H]
        \SetAlgoLined
        \textbf{P1} and \textbf{P2} compute respectively $\alpha_1^2$ and $\beta_2^1$ via \texttt{distr\_product} of $p_1q_2$\\
        \textbf{P1} and \textbf{P2} compute respectively $\beta_1^2$ and $\alpha_2^1$ via \texttt{distr\_product} of $p_2q_1$\\
        \textbf{P1} computes $f_1=p_1q_1+\alpha_1^2+\beta_1^2$\\
        \textbf{P2} computes $f_2=p_2q_2+\alpha_2^1+\beta_2^1$\\
        \textbf{P1} and \textbf{P2} share $f_1$ and $f_2$ to compute $N=f_1+f_2$
        \caption{Distributed computation of $p\cdot q$}
    \end{algorithm}
\end{minipage}
\par
\vspace{5mm}
}

\subsubsection{Biprimality test}
This test, inspired by~\cite{Boneh1997}, is composed of two parts, and requires to perform $s$ times a random filtering test.

{\centering
\begin{minipage}{.8\linewidth}
  \begin{algorithm}[H]
    \SetAlgoLined
    \textbf{P1} samples $\gamma\in\mathbb{Z}_N^\times$ with Jacobi symbol over $N$ equal to 1\\
    \textbf{P1} sends $\gamma$ to \textbf{P2}\\
    \textbf{P1} computes $\gamma_1=\gamma^{\frac{N+1-p_1-q_1}{4}}\mod N$ and sends it to \textbf{P2}\\
    \textbf{P2} computes $\gamma_2=\gamma_1\cdot\gamma^{\frac{-p_2-q_2}{4}}\mod N$\\
    \textbf{if} $\gamma_2=\pm1$ \textbf{P2} sends $\top$ \textbf{else} \textbf{P2} sends $\bot$
    \caption{Distributed filtering biprimality test}
    \end{algorithm}
\end{minipage}
\par
\vspace{5mm}
}

Then the parties verify in a distributed way that $\texttt{gcd}(N, p+q-1) = 1$.

{\centering
\begin{minipage}{.8\linewidth}
  \begin{algorithm}[H]
\SetAlgoLined
\textbf{P1} and \textbf{P2} generate respectively $r_1$ and $r_2$ in $\mathbb{Z}_N$\\
They compute shares $\alpha_1$, $\alpha_2$ of $r_1(p_2+q_2-1)\mod 2^{3k_2}$\\
They compute shares $\beta_1$, $\beta_2$ of $r_2(p_1+q_1)\mod 2^{3k_2}$\\
\textbf{P1} sends $s_1=r_1(p_1+q_1)+\alpha_1+\beta_1\mod N$ to \textbf{P2}\\
\textbf{P2} sends $s_2=r_2(p_2+q_2-1)+\alpha_2+\beta_2\mod N$ to \textbf{P1}\\
Both parties check that $gdc(s_1+s_2,N)=1$ and return $\top$ otherwise return $\bot$
\caption{Distributed test of $\gcd(N,p+1-1)$}
\end{algorithm}
\end{minipage}
\par
\vspace{5mm}
}

\section{Generalization to \textit{n} participants}\label{sec:generalization}
The aforementioned two-party protocol is proven to work securely in a \emph{semi-honest} environment. Here, we will present our generalization to $n$ parties while keeping the same high-level structure. Also, we make sure that the computational complexity and the network overhead are reasonably low, to keep the efficiency of the algorithm similar to the one in~\cite{Frederiksen}.

\subsection{System model}
    We consider a set $\mathcal{N}$ of $n\in\mathcal{N}$ nodes which participates to the protocol. We refer to node $i$ as $\textrm{\textbf{P}}_i$. For the sake of simplicity, we assume the number of participants to be $n = 2^t$, where $t\in\mathbb{N}$.
    
    We keep the assumption of a perfect networking layer and semi-honest participants. However, we also consider the scenario where some of them can collude and communicate their private information in order to derive the private key. From the point of view of an honest node, it is not possible to detect such a behaviour as all nodes actually follow the protocol.
    
\subsection{Key's parts generation}
Each node $\textrm{\textbf{P}}_i$ has to generate a pair $p_i, q_i\in[1,2^k]$. Moreover, we define $p\triangleq\sum\limits_{i=1}^{n}p_i$ and $q\triangleq\sum\limits_{i=1}^{n}q_i$. Since, the biprimality test requires that $p\equiv3(\textrm{mod4})$ and $q\equiv3(\textrm{mod4})$ (see Eq.~\eqref{eq:biprimality_condition}), the protocol enforces that:

\begin{itemize}
    \item $\textrm{\textbf{P}}_1$ randomly picks two numbers $\hat{p}_1$, $\hat{q}_1 \in [1, 2^{k-2}]$ and concatenates two ones to satisfy Eq.~\eqref{eq:biprimality_condition}, i.e., $p_1 \triangleq \hat{p}_1 || \{1;1\}$ and $q_1 \triangleq \hat{q}_1 || \{1;1\}$
    \item all the other nodes $i$, where $i\in\mathcal{N}-\{1\}$, randomly picks two numbers $\hat{p}_i$, $\hat{q}_i \in [1, 2^{k-2}]$, and concatenates two zeros, i.e., $p_i \triangleq \hat{p}_i || \{0;0\}$ and $q_i \triangleq \hat{q}_i || \{0;0\}$.
\end{itemize}

Node $\textrm{\textbf{P}}_1$ can be selected through leader election (potentially based on some node characteristics, such as reputation or hash). We note that the selection of this node does not influence the outcome of the protocol, as it only affects the generation of the two last bits of its secret numbers.

\subsection{Fast trial divisions}
Generalizing the $\beta$-division to $n$ parties is complicated as the original algorithm was specifically designed for two parties. We unstress the constraint of using OTs to communicate the secret shares remainder modulus $\beta$. Our $\beta$-division works in $t$ turns. We set $\mathcal{N}_{0}=\mathcal{N}$ and during the first turn, there is a consensus on a selection of a subset $\mathcal{N}_1\subset\mathcal{N}$ and a bijection $$f_1 \colon \mathcal{N}_{0}\setminus \mathcal{N}_{1} \to \mathcal{N}_1$$ between $\mathcal{N}_1$ and the set of non-selected nodes. Each non-selected node $i$ will send its secret share $p_i\mod \beta$ to its associated node $j$ in $\mathcal{N}_1$ which will sum it with its own secret share modulus $\beta$. In the second turn, we will again select a subset $\mathcal{N}_2\subset\mathcal{N}_1$ of half of the nodes in $\mathcal{N}_1$ and a bijection $$f_2 \colon \mathcal{N}_{1}\setminus \mathcal{N}_{2} \to \mathcal{N}_2$$ between the selected and non-selected nodes such that the non-selected nodes will send the sum of the private share modulus $\beta$ they know. And the algorithm will iteratively divide the selected sets until only $\mathcal{N}_t$ which contains only one node that will be able to output if $\sum\limits_{i\in\mathcal{N}}p_i\mod \beta == 0$. The \autoref{alg:gen_fast} shows a pseudocode of this procedure.

The major issue in this algorithm is how to construct $f_k \colon \mathcal{N}_{k-1}\setminus \mathcal{N}_{k} \to \mathcal{N}_k$ in a decentralized way such that everybody have the same and with a minimum amount of communications. We propose in \autoref{alg:f_k} a way to solve this problem using a deterministic attribution of the associations using a hashing function preventing malicious parties to manipulate this part of the protocol. We assume the parties agree before the fast trials part on a \texttt{seed} and a certain hashing function $H$ such that we can construct $H_m \colon \{0,1\}^* \to [1,m]$ for $m\in\mathbb{N}$.The way this \texttt{seed} and $H$ are generated depends on the network's goals and policies but they can agree to use the \texttt{SHA256} hashing function and generate the \texttt{seed} using a distributed number generator. This technique allows to have no communication for finding consensus during the protocol.

{\centering
\begin{minipage}{.8\linewidth}
  \begin{algorithm}[H]
    \SetAlgoLined
    $\textbf{P}_i.v \leftarrow p_i$, $\forall i\in\mathcal{N}$\\
    \textbf{for} $k \leftarrow 1$\ to\ $k \leftarrow t$\\
    \tab The nodes in $\mathcal{N}_{k-1}$ select $\mathcal{N}_k\subset\mathcal{N}_{k-1}$ containing half the nodes of $\mathcal{N}_{k-1}$\\ \tab and $f_k$\\
    \tab \textbf{foreach} party $\textbf{p} \in \mathcal{N}_{k-1}\setminus \mathcal{N}_{k}$\\
    \tab\tab $\textbf{p}$ sends $\textbf{p}.v \mod \beta$ to $f_k(\textbf{p})$\\
    \tab\tab $f_k(\textbf{p})$ computes  $f_k(\textbf{p}).v \leftarrow (f_k(\textbf{p}.v)+\textbf{p}.v)\mod\beta$\\
    The last party $\textbf{p}_{final}$ in $\mathcal{N}_{t}$ sends $\bot$ if $\textbf{p}_{final}.v=0$ else $\top$
    \caption{Generalization of the fast trial division}\label{alg:gen_fast}
    \end{algorithm}
\end{minipage}
\par
\vspace{5mm}
}

{\centering
\begin{minipage}{.8\linewidth}
  \begin{algorithm}[H]
\SetAlgoLined
$m \leftarrow 2^{t-k-1}$\\
We assume $\mathcal{N}_{k-1}\setminus \mathcal{N}_k$ and $\mathcal{N}_k$ are indexable data structures of $m$ elements\\
$D$ is a data structure mapping  $\mathcal{N}_{k-1}\setminus \mathcal{N}_k\to \mathcal{N}_k$\\
$A\leftarrow[]$\\
\textbf{for} $i\leftarrow1$ to $i\leftarrow m$\\
\tab $j\leftarrow0$\\
\tab \textbf{do}\\
\tab\tab $c\leftarrow H_m(\beta + \texttt{seed} + "|" + k + "|"+ j)$\\
\tab \textbf{while} $c\in A$\\
\tab $D[\mathcal{N}_{k-1}\setminus \mathcal{N}_{k}[i]] \leftarrow \mathcal{N}_{k}[c]$\\
\textbf{return} D\\
\caption{Construction of $f_k$}\label{alg:f_k}
\end{algorithm}
\end{minipage}
\par
\vspace{5mm}
}

The input of the hashing function in this algorithm contains $\beta$, \texttt{seed},$k$ and $j$.

\begin{itemize}
    \item $\beta$ is here to change the output of the function at each division trial to prevent people from getting information from the same parties each time without having to agree on a new seed each time.
    \item $k$ plays the same role be between each turn of one $\beta$-division
    \item $j$ is here to make sure that $A$ is a permutation of $[1,m]$ and does not contain twice the same number while allowing people to compute on their own without communicating.
\end{itemize}

%Let us assume we have $2^t$ parties involved in the computation and we start with each party having $\textrm{candidate}_i = p_i \% \beta$. Then for $t$ steps, we will sample half of the remaining parties and we will map in a bijective way each selected party to a non-selected one. Then for all the non-selected $A_i$, they will transmit $\textrm{candidate}_i$ to their selected neighbour $A_j$ which will compute $\textrm{candidate}_j = (\textrm{candidate}_j+\textrm{candidate}_i) \% \beta$.

%We suggest all the parties agree on a certain seed and a hashing function such that the mapping cannot be guessed but can be verified quickly to prevent attacks. As the selected primes should be tiny compared to the size of $p_i$ (around $2^{256}$), we believe we don't share much information when it comes to sharing in the very first round although it might be a major security breach in this protocol. Besides, as we change the seed at each divisibility trial, it makes it harder to infer $p_i$ from the shared rests.

%One security issue could be that if the party that makes the verification is malicious and decides not to declare it's divisible by $\beta$. However it is easy once the public key finally generated to run on a single computer fast trial divisions to ensure this part hasn't been compromised. Thus we suggest people to run divisions once the public key is generated.

\subsection{Distributed multiplication}
In this subsection, we want to compute $\sum_i p_i \cdot \sum_i q_i$ in a similar way to the Frederiksen's algorithm. First, note that
\begin{equation}\label{eq:distr_mul}
    \sum\limits_{i=1}^{n} p_i \cdot \sum\limits_{i=1}^{n} q_i = \sum\limits_{i=1}^{n} (p_i\cdot q_i) + \sum\limits_{i=1}^{n-1}\sum\limits_{j=i+1}^{n}(p_i\cdot q_j+p_j\cdot q_i)
\end{equation}

The first term of the left hand side of \autoref{eq:distr_mul} can be computed solely by every party. As for the second term, for each pair ${i,j}\in\mathcal{N}^2$ with $i<j$, the nodes $\textrm{\textbf{P}}_i$ and $\textrm{\textbf{P}}_j$ have to compute $p_i\cdot q_j + p_j\cdot q_i$ in a distributed way without disclosing their secret shares. We then suggest that $\textrm{\textbf{P}}_i$ and $\textrm{\textbf{P}}_j$ run the \texttt{distr\_product} routine and compute $x_i^{j,p}$ for $\textrm{\textbf{P}}_i$ and $x_j^{i,q}$ for $\textrm{\textbf{P}}_i$ for computing $p_i\cdot q_j$ and respectively they would compute $x_i^{j,q}$ and $x_j^{i,p}$ while computing $p_j\cdot q_i$.

Finally, we have for each party $\textrm{\textbf{P}}_i$, $\textrm{\textbf{P}}_i$ computes 
\begin{equation*}
    f_i = p_i\cdot q_i + \sum\limits_{j\in\mathcal{N}, j\neq i} x_i^{j,p} + x_i^{j,q},
\end{equation*}
and broadcast it to everyone so that everybody can compute $N = \sum\limits_{i=1}^{n} f_i$. At that point, N is a public information to everyone.

{\centering
\begin{minipage}{.8\linewidth}
  \begin{algorithm}[H]
\SetAlgoLined
$f_i\leftarrow 0$ for $i\in\mathcal{N}$\\
\textbf{for} $1\leq i\leq j\leq n$\\
\tab \textbf{if} $i=j$\\
\tab\tab $f_i \pluseq p_iq_i$\\
\tab \textbf{else}\\
\tab\tab We compute $p_iq_j$ and $p_jq_i$ as in \autoref{alg:alpha_beta}\\
\tab\tab $f_i \pluseq x_i^{j, p} + x_i^{j, q}$\\
\tab\tab $f_j \pluseq x_j^{i, p} + x_j^{i, q}$\\
The parties share $f_i\ \forall i\mathcal{N}$ to compute $N\leftarrow\sum\limits_{i\in\mathcal{N}} f_i$\\
\caption{Generalized distributed multiplication}
\end{algorithm}
\end{minipage}
\par
\vspace{5mm}
}

{\centering
\begin{minipage}{.8\linewidth}
  \begin{algorithm}[H]
\SetAlgoLined
We have two parties, \textbf{P$i$} and \textbf{P$j$} with $i<j$\\
\textbf{P$i$} and \textbf{P$j$} compute respectively $x_i^{j, p}$ and $x_j^{i, q}$ via distributed computation of $p_iq_j$\\
\textbf{P$i$} and \textbf{P$j$} compute respectively $x_i^{j, q}$ and $x_j^{i, p}$ via distributed computation of $p_jq_i$\\
\caption{Computing $p_iq_j+p_jq_i$}\label{alg:alpha_beta}
\end{algorithm}
\end{minipage}
\par
\vspace{5mm}
}

\subsection{Biprimality test}

\subsubsection*{First part}
In this section we keep the same idea of testing for a certain $\gamma$ randomly generated if $\gamma^{N+1-p-q}=\pm1$. This part can be easily done in a decentralized way with properties offered by the exponentiation. The parties elect a leader $\textrm{\textbf{P}}_e$ that will generate a number $\gamma\in\mathbb{Z}_N^\times$ with Jacobi symbol over $N$ equal to $1$ and shares it with everybody. Then each party $\textrm{\textbf{P}}_i, i\in\mathcal{N}$ will compute $\gamma_i = \gamma^{-p_i-q_i}$ and send it to $\textrm{\textbf{P}}_e$. Once $\textrm{\textbf{P}}_e$ received all the $gamma_i$, it will compute $
gamma{final} = \gamma^{N+1}\cdot\prod\limits_{i=1}^{n}\gamma_i\mod N$. Finally, if $\card{\gamma_{final}} = 1$, $\textrm{\textbf{P}}_e$ broadcasts $\top$ otherwise it broadcasts $\bot$. The \autoref{alg:gen_first_biprim} gives the details of this procedure. This test returns $\top$ if $N$ is composite of two prime numbers but have a probability of $1/2$ to return $\top$ otherwise. Therefore, in order to increase the reliability of this algorithm, the parties should run it $s$ time with $s$ such that $\frac{1}{2^n}$ is a probability low enough to return $\top$ for a wrong modulus.
 
{\centering
\begin{minipage}{.8\linewidth}
  \begin{algorithm}[H]
\SetAlgoLined
Parties elect a party $\textrm{\textbf{P}}_e$\\
\textbf{P$_s$} generates $\gamma\in\mathbb{Z}_N^\times$ and shares with everybody\\
\textbf{for} party \textbf{P$_i$} (including $\textrm{\textbf{P}}_e$)\\
\tab \textbf{P$_i$} computes $\gamma_i=\gamma^{-p_i-q_i}$\\
\tab \textbf{P$_i$} sends $\gamma_i$ to $\textrm{\textbf{P}}_e$\\
$\textrm{\textbf{P}}_e$ compute $\gamma_{final}=\gamma^{N+1}\cdot\prod\limits_{i=1}^{n}\gamma_i \mod N$\\
\textbf{if} $\card{\gamma_{final}} = 1$, $\textrm{\textbf{P}}_e$ broadcasts $\top$ else $\bot$\\
\caption{First discriminating biprimalty test}\label{alg:gen_first_biprim}
\end{algorithm}
\end{minipage}
\par
\vspace{5mm}
}
 
We can see it is important for $\gamma$ to be in $\mathbb{Z}_N^\times$ in order to be able to compute negative exponents by using $\gamma^{-1}$. The way of electing $\textrm{\textbf{P}}_e$ . A way that can be done is using the hashing function and the seed defined in the fast trial divisions in order to elect implicitly the leading party at each trial without requiring further communications.

\subsubsection*{Second part}
In this part, we want compute $\texttt{gcd}(\sum_{i\in\mathcal{N}} r_i\cdot(\sum_{i\in\mathcal{N}} p_i + \sum_{i\in\mathcal{N}} q_i - 1), N)$ where $r_i$ is a number generated by a node $i\in\mathcal{N}$. Let $\Delta_i=p_i+q_i, i\in\mathcal{N}$. In the same idea as for the distributed multiplication, we can rewrite $\sum_i\in\mathcal{N} r_i\cdot(\sum_i\in\mathcal{N} p_i + \sum_i\in\mathcal{N} q_i - 1)$ as 

\begin{equation}\label{eq:gcd_distr}
\sum\limits_{i\in\mathcal{N}} r_i\cdot(\sum\limits_{i\in\mathcal{N}} p_i + \sum\limits_{i\in\mathcal{N}} q_i - 1) = \sum\limits_{i\in\mathcal{N}} (r_i\cdot\Delta_i - 1) + \sum\limits_{i\in\mathcal{N}} \sum\limits_{\substack{i\in\mathcal{N} \\ i\neq j}} r_i\cdot\Delta_i
\end{equation}

The have the same schema as for the distributed multiplication whith the first term of the left hand of \autoref{eq:gcd_distr} which can be computed solely by a node and the other term has to be computed using \texttt{distr\_product}. Finally they share all the sum $g_i$ of their parts of computation similarly to $f_i$ in the distributed multiplication and they compute $G=\sum_{i\in\mathcal{N}} g_i\mod N$. Finally each node can verify on its own wether $\gcd(G, N) = 1$. If so, the modulus $N$ can be used, otherwise, the protocol has to be restarted.

{\centering
\begin{minipage}{.8\linewidth}
  \begin{algorithm}[H]
\SetAlgoLined
$g_i\leftarrow 0$ for $i\in\mathcal{N}$\\
Each party $i$ generate a secret random number $r_i\in\mathbb{Z}_N$ for $i\in\mathcal{N}$\\
\textbf{for} $1\leq i\leq j\leq n$\\
\tab \textbf{if} $i=j$\\
\tab\tab $s_i \pluseq r_i(\Delta_i-1)$\\
\tab \textbf{else}\\
\tab\tab We compute $r_i\Delta_j$ and $r_j\Delta_i$ as in \autoref{alg:alpha_beta}\\
\tab\tab $g_i \pluseq u_i^{j,r}+ u_i^{j,\Delta}$\\
\tab\tab $g_j \pluseq u_j^{i,\Delta}+\beta_j^{i,r}$\\
The parties share $g_i\mod N\ \forall i\in\mathcal{N}$ to compute $G=\sum\limits_{i=1}^{n} g_i\mod N$\\
\textbf{if} $\texttt{gcd}(G, N)=1$ then $N$ is considered to be sure\\
\caption{Second discriminating biprimality test}
\end{algorithm}
\end{minipage}
\par
\vspace{5mm}
}

\section{Network overhead}\label{sec:analysis}

%\subsection{Amount of communications}
Here we present a theoretical analysis of the amount of communications necessary for a successful try. We use \textbf{br} as an abbreviation for ``broadcast''.

\subsection{Fast trial divisions}
Considering a single division, if we consider for simplicity we have $n=2^t$ participants, we have $t$ steps where for the $i$-th step, $2^{t-i}$ parties send their values to someone else. In the worst case, which is the node which will return the value, it makes $\log_2(n)$ communications to receive the data plus one communication to return the result. In the best case, which is a node communicating it's private share modulus the small prime at the first round, it only makes on connection to give and one to receive the output.

So we have for $B$ different fast trial divisions, each party operates the following amount of communications:

\begin{itemize}
    \item \textbf{Worst-Case :} $B\cdot(\log_2(n)+1)$
    \item \textbf{Best-Case :} $2\cdot B$
\end{itemize}

\subsection{Distributed multiplication}
Let us count for a single multiplication of $a\cdot b$ with two $k$-bits numbers. For each bit the two parties instantiate an OT and either input or receives a value from it which makes $k$ communications per party. Then as each party computes two product with the other parties, they operate $2k(n-1)$ communications plus $2k(n-1)$ OT instantiations. Finally we have a question of how sharing the final values in order to add everything to get $N$. We can suppose everyone broadcasts their final value $f_i$.

Finally we get $2k(n-1) + n$ communications per party for the distributed multiplication plus $2k(n-1)$ oblivious transfer initializations. 

\subsection{Biprimality test - First part}
In this part, the chosen party to generate $\gamma$ only broadcasts the value of $\gamma$, then the other parties communicate their exponentiation with a single communication and finally the first party communicates the output with a broadcast.

Then we have for each party according to the role during the $s$ trials

\begin{itemize}
    \item \textbf{Worst-Case :} $s\cdot(1+(n-1)+1) = s\cdot(n+1)$
    \item \textbf{Best-Case :} $3\cdot s$
\end{itemize}

\subsection{Biprimality test - Second part}
Here we have something really similar to the distributed multiplication part except the multiplications ar $2k$ bits longs here. Then each party operates $4k(n-1)+n$ communications plus $4k(n-1)$ OT instantiations.

We can see the theoretical amount of communication is linear with the growth of $n$ and $k$ so we have a quadratic growth of the global amount of communications. Furthermore, we can parallelize a great part of the steps. For example we can run various fast trial divisions in parallel with a different prime number and massively parallelize the multiplications.

%\subsection{Numbers distribution}
%When summing numbers sampled in a uniform distribution, it leads to a normal law. Indeed, the variance 

%\begin{figure}[ht] 
%  \label{ fig7} 
%  \begin{minipage}[b]{0.5\linewidth}
%    \centering
%    \includegraphics[width=\linewidth]{images/1.png} 
%    \caption{Distribution for 1 number} 
%    \vspace{4ex}
%  \end{minipage}%%
%  \begin{minipage}[b]{0.5\linewidth}
%    \centering
%    \includegraphics[width=\linewidth]{images/2.png} 
%    \caption{Distribution for 2 number} 
%    \vspace{4ex}
%  \end{minipage} 
%  \begin{minipage}[b]{0.5\linewidth}
%    \centering
%    \includegraphics[width=\linewidth]{images/50.png} 
%    \caption{Distribution for 50 number} 
%    \vspace{4ex}
%  \end{minipage}%% 
%  \begin{minipage}[b]{0.5\linewidth}
%    \centering
%    \includegraphics[width=\linewidth]{images/500.png} 
%    \caption{Distribution for 500 number} 
%    \vspace{4ex}
%  \end{minipage} 
%\end{figure}

%\includegraphics{pdfs/rsa_generation.pdf}
\printbibliography

\appendix

\section*{$1$-out-of-$n$ Oblivious Transfer}\label{appendix:ot}
A $1$-out-of-$n$ Oblivious Transfer (OT) is a cryptosystem which allows a sender to propose $n$ messages and a receiver to chose one of them without the sender knowing which one. We can also consider a random 1-out-of-$n$ OT which sends the $n$ messages to the sender instead of receiving an input. This cryptosystem allows two people to share information without disclosing knowledge on the choice of the receiver and the other options he could have chosen. This will be used as a way to compute distributed multiplications and in other situations in the protocol. The behaviour of this mechanism is specified in \autoref{alg:ot}.

\begin{algorithm}[H]
\SetAlgoLined
\textbf{P1} and \textbf{P2} initialize the OT with the parameter $n$\\
\textbf{P1} sends $\{\mu_1,\dots,\mu_n\}$\\
\textbf{P2} sends $i\in[1,n]$\\
\textbf{P2} receives $\mu_i$\\
\caption{$1$-out-of-$n$ Oblivious Transfer (OT) procedure}\label{alg:ot}
\end{algorithm}

\end{document}